\documentclass[a4paper,fleqn]{cas-dc}

\usepackage{amssymb}
\usepackage{amsmath}
\usepackage{bm}
\usepackage{color}
\usepackage{epsfig}
\usepackage{multirow}
\usepackage{multicol}
\usepackage[numbers]{natbib}
\usepackage{subfigure}
\usepackage{threeparttable}
\usepackage[switch]{lineno}

\newcommand{\s}{$\sqrt{s}$ }
\newcommand{\sNN}{$\sqrt{s_{\mathrm{NN}}}$ }
\newcommand{\GeVc}{GeV/$c$}

\newcommand{\pT}{$p_{\mathrm{T}}$}
\newcommand{\dz}{$\mathrm{D^{0}}$}
\newcommand{\dpm}{$\mathrm{D^{\pm}}$}
\newcommand{\ds}{$\mathrm{D_{s}}$}
\newcommand{\lc}{$\mathrm{\Lambda_{c}}$}
\newcommand{\jpsi}{$\mathrm{J/\psi}$}
\newcommand{\RAA}{R_{\mathrm{AA}}}
\newcommand{\fPb}{f_{\mathrm{PbPb}}}
\newcommand{\fpp}{f_{\mathrm{pp}}}
\newcommand{\cte}{\mathrm{c\rightarrow e}}
\newcommand{\bte}{\mathrm{b\rightarrow e}}
\newcommand{\Dte}{\mathrm{D\rightarrow e}}
\newcommand{\Lcte}{\mathrm{\Lambda_c\rightarrow e}}
\newcommand{\jpsite}{\mathrm{J/\psi\rightarrow e}}
\newcommand{\jpsitee}{$\mathrm{J/\psi\rightarrow e^+e^-}$}
\graphicspath{{figs/}}

\def\tsc#1{\csdef{#1}{\textsc{\lowercase{#1}}\xspace}}
\tsc{WGM}
\tsc{QE}
\tsc{EP}
\tsc{PMS}
\tsc{BEC}
\tsc{DE}

\begin{document}

\let\WriteBookmarks\relax
\def\floatpagepagefraction{1}
\def\textpagefraction{.001}
\shorttitle{Charm and beauty isolation from heavy flavor decay electrons in p+p and Pb+Pb collisions at $\sqrt{s_{\mathrm{NN}}}$ = 5.02 TeV at LHC}
\shortauthors{Dongsheng Li et~al.}

\title [mode = title]{Charm and beauty isolation from heavy flavor decay electrons in p+p and Pb+Pb collisions at $\sqrt{s_{\mathrm{NN}}}$ = 5.02 TeV at LHC}

\author[1]{Dongsheng Li}
\author[1]{Fan Si}
\author[1]{Yidan Zhao}
\author[1]{Pengyu Zhou}
\author[1]{Yifei Zhang}\cormark[1]\ead{ephy@ustc.edu.cn}
\author[1]{Xiujun Li}
\author[1]{Chengxi Yang}
\address[1]{State Key Laboratory of Particle Detection and Electronics, University of Science and Technology of China, Hefei 230026, China}

\cortext[cor1]{Corresponding author}

\begin{abstract}
We present an analysis on the heavy flavor hadron decay electrons with charm and beauty contributions decomposed via a data driven method in p+p and Pb+Pb collisions at $\sqrt{s_{\mathrm{NN}}}$ = 5.02 TeV at LHC. The transverse momentum $p_{\mathrm{T}}$ spectra, nuclear modification factor $R_{\mathrm{AA}}$ and azimuthal anisotropic flow $v_2$ distributions of electrons from charm and beauty decays are obtained. We find that the electron $R_{\mathrm{AA}}$ from charm ($R_{\mathrm{AA}}^{\mathrm{c\rightarrow e}}$) and beauty ($R_{\mathrm{AA}}^{\mathrm{b\rightarrow e}}$) decays are suppressed at $p_{\mathrm{T}}$ $>$ 2.0 \GeVc\ in Pb+Pb collisions, which indicates that charm and beauty interact with and lose their energy in the hot-dense medium. A less suppression of $R_{\mathrm{AA}}^{\mathrm{b\rightarrow e}}$ than $R_{\mathrm{AA}}^{\mathrm{c\rightarrow e}}$ at 2.0 $<$ $p_{\mathrm{T}}$ $<$ 8.0 GeV/$c$ is observed, which is consistent with the mass-dependent partonic energy loss scenario. A non-zero electron $v_2$ from beauty decays ($v_{2}^{\mathrm{b\rightarrow e}}$) is observed and in good agreement with ALICE measurement. At low $p_{\mathrm{T}}$ from 1.0 to 3.0 GeV/$c$, a discrepancy between RHIC and LHC results is observed with an 85\% confidence level, which might be a possible hint of an energy-dependent behavior of beauty quarks reacting with the medium created in heavy-ion collisions. At 3.0 $<$ $p_{\mathrm{T}}$ $<$ 7.0 GeV/$c$, $v_{2}^{\mathrm{b\rightarrow e}}$ deviates from a number-of-constituent-quark (NCQ) scaling hypothesis, which favors that beauty quark is unlikely thermalized in heavy-ion collisions at LHC energy.
\end{abstract}

\begin{keywords}
Quark-Gluon Plasma \sep charm \sep beauty \sep semileptonic decay \sep nuclear modification factor \sep elliptic flow
\end{keywords}

\maketitle

\section{Introduction}

Strong interactions among quarks and gluons by exchanging color charges are described by the theory of Quantum Chromodynamics (QCD). Quarks and gluons are confined in color-neutral hadrons and can be liberated under an extremely hot and dense condition such as the stage of a few microseconds after the big bang of the universe. Ultra-relativistic heavy-ion collisions provide such environment to create the state of matter known as quark-gluon plasma (QGP) that behaves almost as a perfect fluid~\cite{Gyulassy:2004vg, Bass:1998vz, Karsch:1995sy}. Heavy quarks (charm and beauty) are produced mainly via initial hard scatterings in heavy-ion collisions due to their large masses. Their production yield can be evaluated by perturbative QCD. Thus, heavy quarks are believed to experience the whole medium dynamic evolution after their creation and maintain conserved quark numbers. Therefore, heavy quarks are an ideal probe to study QGP matter and sensitive to its thermal dynamic properties~\cite{Moore:2004tg,Tang:2020ame,Luo:2020pef}. In the past ten years, open heavy flavor hadrons have been measured in many experiments at Relativistic Heavy Ion Collider (RHIC) and Large Hadron Collider (LHC)~\cite{STAR:2005gfr, PHENIX:2004vcz, Muller:2012zq, Akiba:2015jwa}. It was found that the nuclear modification factors ($\RAA$) of both electrons from heavy flavor decays and open charm mesons were strongly suppressed at high transverse momentum (\pT) in central heavy-ion collisions. This indicates that charm quarks significantly lose their energy when traversing the hot-dense medium. A significant elliptic flow ($v_2$) of open charm mesons is also observed and comparable with light flavor hadrons, which indicates that charm quark reaches thermalization in heavy-ion collisions.

Beauty, with a factor of three larger mass than charm, is predicted to lose less energy than charm in the medium created in heavy-ion collisions either with a suppression of the gluon radiation angle~\cite{Dokshitzer:2001zm, Djordjevic:2004nq, Buzzatti:2011vt} or the elastic collisions~\cite{Moore:2004tg,Cao:2017hhk}. And its $v_2$ is probably different from those of charm quark and light flavors if it is far from thermalization. However, measurement of beauty is challenging due to its small production cross section and low hadronic decay branching ratios. Thus, in an alternative way, the beauty was commonly measured indirectly via its decay products, such as electrons, non-prompt \dz\ and \jpsi\ mesons. It was observed that the $\RAA$ of its decay daughters are less suppressed compared to those of prompt open charm mesons at \pT\ up to $\sim$ 20 GeV/$c$, which is consistent with the mass hierarchy of the parton energy loss in the medium~\cite{STAR:2014wif, STAR:2018zdy, STAR:2006btx, PHENIX:2006iih, ALICE:2021rxa, CMS:2017uuv, ATLAS:2020yxw, ATLAS:2021xtw}. Recently, a sizable electron $v_2$ from beauty decays was also measured by ALICE~\cite{ALICE:2020hdw}, which indicates strong interactions between beauty and the medium. Since charm looks no significant difference compared with light quarks in terms of $\RAA$ and $v_2$ in RHIC and LHC energies~\cite{Tang:2020ame,ALICE:2020iug}, it is crucial to study whether beauty is sensitive to the different media created at RHIC and LHC energies and behaves differently under different temperatures. 

In a previous study, the charm and beauty contributions are separated from the inclusive heavy flavor electrons (HFE) in 200 GeV Au+Au collisions at RHIC~\cite{Si:2019esg}. It was found that the electron $\RAA$ from beauty decays is less suppressed compared to that from charm decays, which is consistent with the mass-dependent energy loss expectation. And a first zero electron $v_2$ from beauty decays, lower than a number-of-constituent-quark (NCQ) scaling hypothesis, was observed at low \pT, which indicates that beauty quark is unlikely thermalized and too heavy to be moved following the collective flow of lighter partons in heavy-ion collisions at RHIC energy.

In this paper, we perform an analysis in a similar way to separate charm and beauty contributions from heavy flavor decay electrons in 5.02 TeV p+p and Pb+Pb collisions. The $\RAA$ and $v_2$ of electrons from charm and beauty decays in 5.02 TeV Pb+Pb collisions at LHC are obtained and compared with those in 200 GeV Au+Au collisions at RHIC.

\section{Analysis Technique and Results}

\subsection{Transverse Momentum Spectra and $\RAA$}
\label{sec2.1}

\begin{figure}[!htb]
\centering
\includegraphics[width=\linewidth]{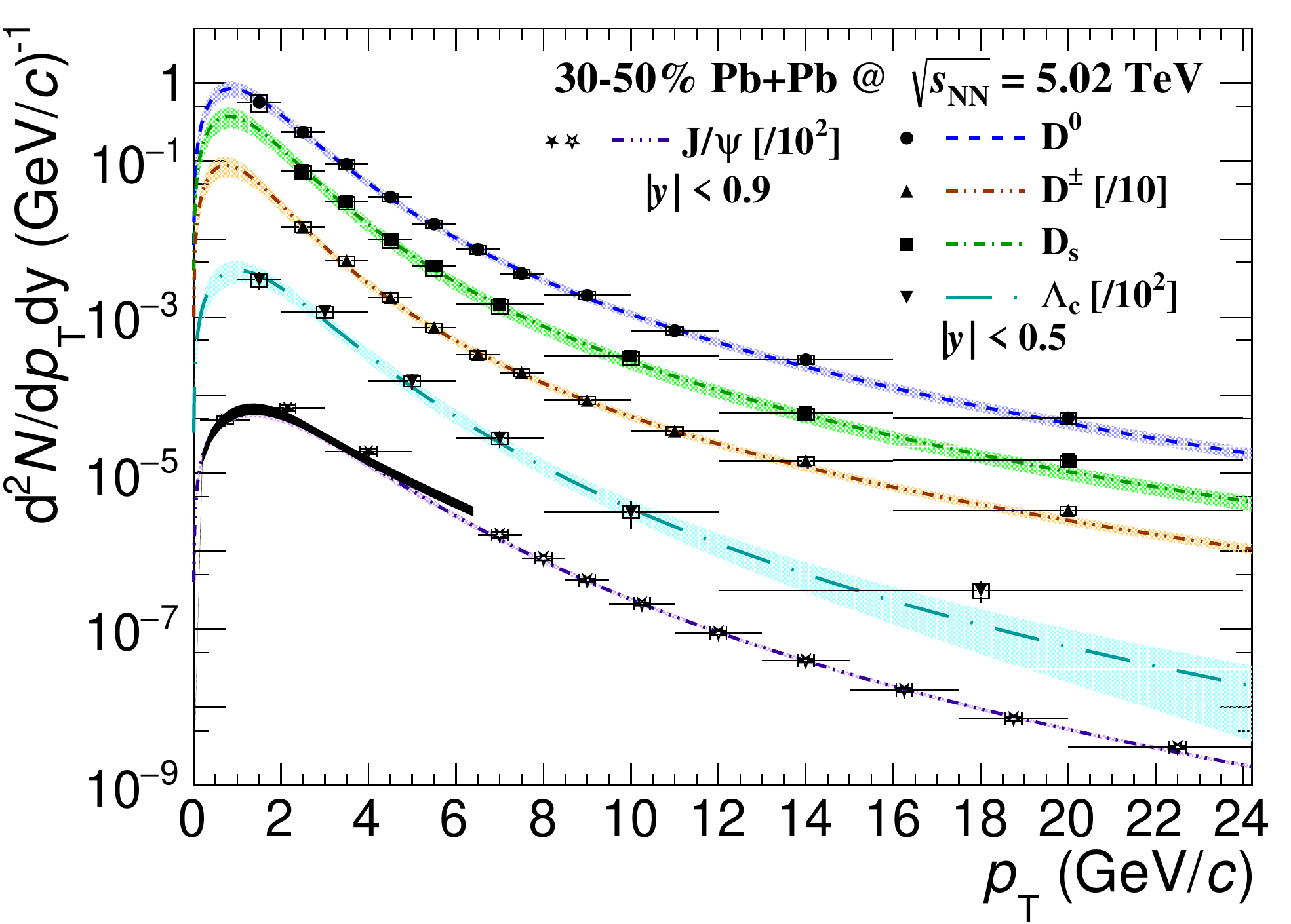}
\vspace{-2ex}
\caption{Parameterized charmed hadron (\dz, \dpm, \ds, \lc, \jpsi) spectra at mid-rapidity ($|y|$ $<$ 0.9 for \jpsi\ and 0.5 for others) in 30--50\% central Pb+Pb collisions at \sNN = 5.02 TeV. Uncertainties are shown as shaded color bands. The experimental measurements and the TAMU prediction are shown as black points and black bands, respectively. The \jpsi\ measurement in 20--40\% centrality class by ALICE~\cite{ALICE:2019nrq} scaled to 30--50\% by $\left\langle N_{\mathrm{coll}}\right\rangle$~\cite{CentNoteAlice} is also shown for comparison (open black stars) but not included in the analysis. The measurement on prompt \jpsi\ in 0--100\% centrality class by CMS~\cite{CMS:2017uuv} scaled to 30--50\% by $\left\langle N_{\mathrm{coll}}\right\rangle$~\cite{CMS:2016xef} (solid black stars) is combined with the TAMU prediction on inclusive \jpsi\ production in the parameterization. The spectra of charmed hadrons are scaled with different factors to separate the merged points.}
\label{fig1}
\end{figure}

We adopt the data-driven method developed in Ref.~\cite{Si:2019esg} to separate charm and beauty decay contributions from the inclusive HFE spectra~\cite{ALICE:2019nuy} in 0--10\% and 30--50\% central Pb+Pb collisions at \sNN = 5.02 TeV and p+p collisions at \s = 5.02 TeV at LHC by taking advantage of the most recent charmed hadron measurements. Different from the previous study at RHIC energy, the HFE measurement~\cite{ALICE:2019nuy} from the ALICE experiment was not subtracted by the \jpsi\ contribution (about 5\% at 2 < \pT\ < 3 \GeVc\ and much lower in other \pT\ regions), and thus, the \jpsitee\ contribution is also studied in this work. In Pb+Pb collisions, the ALICE experiment has precisely measured the \pT\ spectra of prompt \dz, \dpm\ and \ds\ mesons (0 $<$ \pT\ $<$ 50 \GeVc) and \lc\ baryons (0 $<$ \pT\ $<$ 24 \GeVc) including their antiparticles at mid-rapidity ($|y|$ $<$ 0.5) at 0--10\% and 30--50\% centrality classes~\cite{ALICE:2018lyv,ALICE:2021kfc,ALICE:2021bib}. The \jpsi\ spectra ($|y|$ $<$ 0.9) in both centrality classes are provided by the TAMU model predictions~\cite{Zhao:2010nk,Zhao:2011cv,Du:2015wha} combined with the measurement by the CMS experiment~\cite{CMS:2017uuv} at 6.5 $<$ \pT\ $<$ 50 \GeVc, which is scaled from 0--100\% centrality with the average number of binary nucleon-nucleon collisions ($\left\langle N_{\mathrm{coll}}\right\rangle$)~\cite{CMS:2016xef}. All the spectra are then parameterized with a Levy~\cite{Wilk:1999dr} function, and the uncertainties are given by a quadratic sum of two components: (a) 1-$\sigma$ band from the parameterization with statistical uncertainties and (b) systematic uncertainties from the parameterized spectra scaled to upper and lower limits. The parameterization at 30--50\% centrality is shown as an example in Fig.~\ref{fig1}, which works well compared with the experimental data (black points) and the TAMU prediction (black band). In p+p collisions, the prmopt D-mesons ($|y|$ $<$ 0.5)~\cite{ALICE:2021mgk}, \lc\ ($|y|$ $<$ 0.5)~\cite{ALICE:2020wfu} and \jpsi\ ($|y|$ $<$ 0.9)~\cite{ALICE:2021edd} spectra measured by ALICE are parameterized with the same steps as those in Pb+Pb collisions. To extend the \pT\ ranges, the measurements of \jpsi\ by the ATLAS experiment~\cite{ATLAS:2017prf} at 12 $<$ \pT\ $<$ 40 \GeVc\ and \lc\ by CMS~\cite{CMS:2019uws} at 10 $<$ \pT\ $<$ 20 \GeVc\ are combined with the ALICE measurements for parameterization.

\begin{figure}[!htb]
\centering
\includegraphics[width=\linewidth]{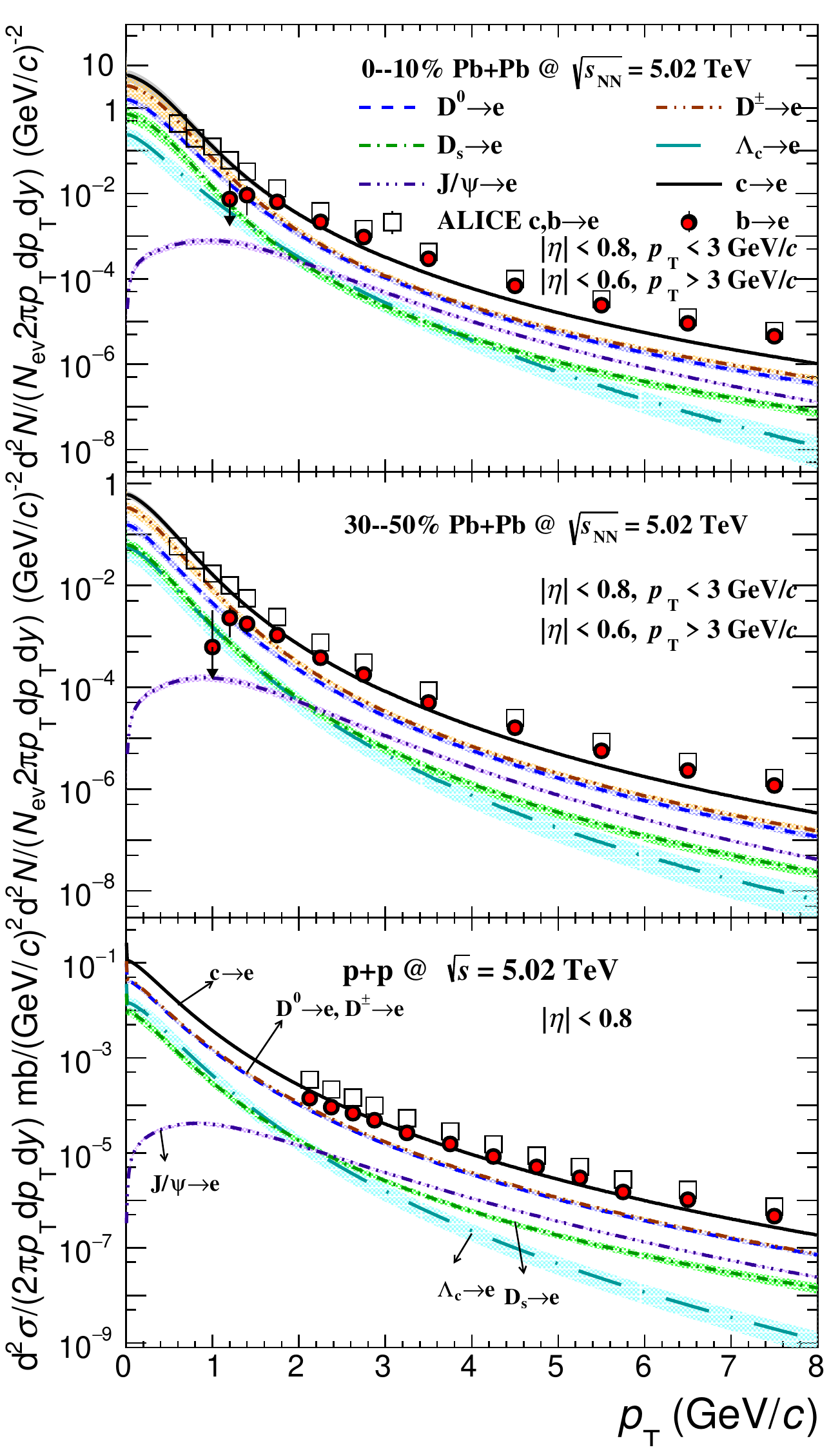}
\vspace{-2ex}
\caption{Electron spectra from charmed hadrons (\dz, \dpm, \ds, \lc, \jpsi) and the sum of them ($\cte$) and the inclusive HFE spectra at mid-rapidity ($|y|$ $<$ 0.8 at \pT\ $<$ 3 \GeVc\ and $|y|$ $<$ 0.6 at \pT\ $>$ 3 \GeVc) in 0--10\% and 30--50\% central Pb+Pb collisions at \sNN = 5.02 TeV and p+p collisions($|y|$ $<$ 0.8) at \s = 5.02 TeV. The spectra of beauty hadron decay electrons ($\bte$) are obtained by subtracting the $\cte$ contributions from the HFE data from ALICE~\cite{ALICE:2019nuy}. Uncertainties are shown as shaded bands.}
\label{fig2}
\end{figure}

\begin{figure}[!htb]
\centering
\includegraphics[width=\linewidth]{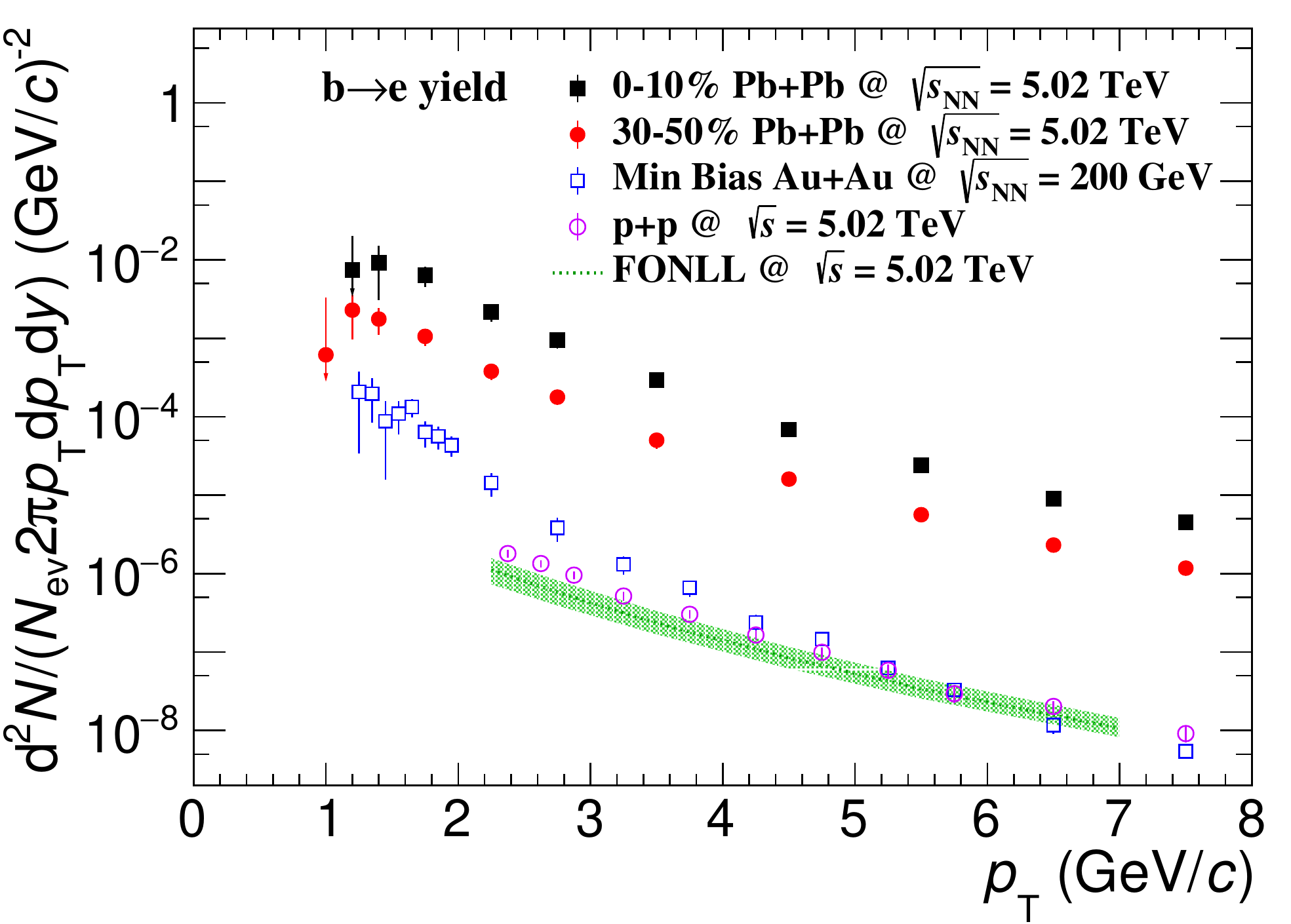}
\vspace{-2ex}
\caption{Electron spectra from beauty hadron decays in 0--10\% and 30--50\% central Pb+Pb collisions at \sNN = 5.02 TeV, p+p collisions at \s = 5.02 TeV compared with the FONLL prediction~\cite{Cacciari:1998it} and the result in minimum bias Au+Au collisions at \sNN = 200 GeV~\cite{Si:2019esg}.}
\label{fig3}
\end{figure}

The charmed hadrons described above are then simulated to decay to electrons via semileptonic decay ($\mathrm{e^+e^-}$ for \jpsi) channels with their parameterized \pT\ spectra at 0 $<$ \pT\ $<$ 20 \GeVc\ and Gaussian rapidity distributions ($\mu$ = 0 and $\sigma$ = 1.7 for open charm hadrons, $\mu$ = 0 and $\sigma$ = 1.9 for \jpsi) as inputs. These rapidity distributions have been checked by the PYTHIA~\cite{Sjostrand:2006za} event generator. By varying the rapidity distributions with a sigma of 1.7 $\pm$ 0.3, the obtained electron spectra changes $<$ 1\% in 0 $<$ \pT\ $<$ 20 GeV/$c$. The open charm hadron decays are conducted with the help of decay formfactors in the hadron rest frame which are sampled from the measured distribution by the CLEO experiment~\cite{Liu:2006id}, while the \jpsi\ decays are simulated with PYTHIA. Note that a non-zero \jpsi\ polarization could have impact on the decay kinematics. Recently the ALICE experiment measured \jpsi\ polarization which shows some hint of non-zero polarization but with large uncertainties~\cite{ALICE:2020iev}. Thus here the effect of potential \jpsi\ polarization on the decay electron spectra is not taken into account. Figure~\ref{fig2} shows the electron spectra from \dz\ (blue dashed curve), \dpm\ (brown dot-dot-dashed curve), \ds\ (green dot-dashed curve), \lc\ (cyan long-dot-dashed curve) and \jpsi\ (violet dot-dot-dot-dashed curve) decays and the summed charm contributions ($\cte$, black solid curve) at mid-rapidity ($|y|$ $<$ 0.8 at \pT\ $<$ 3 \GeVc\ and $|y|$ $<$ 0.6 at \pT\ $>$ 3 \GeVc) in 5.02 TeV 0--10\% and 30--50\% central Pb+Pb collisions and p+p collisions. All components of the total $\cte$ production from individual charmed hadron decays are summarized in Table~\ref{tab1}. The maximum \pT\ of charmed hadron measurements and the contributions of unsimulated regions of charmed hadron decays ($p_{\mathrm{T}}^{\mathrm{charm}}$ $>$ 20 \GeVc) to the electrons in 0--8 \GeVc\ are summarized in Table~\ref{tab4}. The contributions are found to be less than 0.04\% to insure that most electrons in the region of interest come from the simulated region of charmed hadrons. The uncertainties of the charmed hadron \pT\ inputs are propagated into the decay electron spectra, and the uncertainties of branching ratios~\cite{ParticleDataGroup:2018ovx} are also taken into account. The total uncertainties of electron spectra from individual charmed hadron decays are shown as shaded bands in Fig.~\ref{fig2}. All contributions of the total uncertainties of $\cte$ within 0 $<$ \pT\ $<$ 8 \GeVc\ are summarized in Table~\ref{tab2}. The total uncertainties of $\cte$, quadratically summed from all uncorrelated components, are 8.0\%--40.6\% and 6.4\%--17.9\% at 0--10\% and 30--50\% centrality classes in Pb+Pb collisions, and 4.2\%--8.2\% in p+p collisions, respectively. The electron spectra from beauty hadron decays ($\bte$, red solid circles), are then calculated by subtracting the $\cte$ contribution from the inclusive HFE spectra (black open squares)~\cite{ALICE:2019nuy}. Figure~\ref{fig3} presents the $\bte$ spectra in 0--10\% (black solid squares) and 30--50\% (red solid circles) central Pb+Pb collisions at \sNN = 5.02 TeV and in p+p collisions at \s = 5.02 TeV with the previously obtained result (blue open squares)~\cite{Si:2019esg} in minimum bias Au+Au collisions at \sNN = 200 GeV and the FONLL prediction (green open crosses)~\cite{Cacciari:1998it} at \s = 5.02 TeV shown for comparison. The p+p result obtained in this work is consistent with FONLL upper limit. And note that the \pT-differential yield spectra in p+p collisions is obtained by dividing the \pT-differential cross section spectra in Fig.~\ref{fig3} by the minimum bias trigger cross section (50.9 $\pm$ 0.9 mb) at \s = 5.02 TeV. In all of the following uncertainty calculations, statistical and systematic uncertainties are quadratically summed into total uncertainties, unless otherwise specified.

\begin{table}
\centering
\caption{Production components of the $\cte$ spectra (0--8 \GeVc) from individual charmed hadron decays.}
\label{tab1}
\begin{threeparttable}
\resizebox{\linewidth}{!}{ 
\begin{tabular}{cccc}
\multirow{3}{*}{} \\ \hline
centrality & 0--10\% & 30--50\% & p+p\\  \hline
$\mathrm{D^{0}\ \rightarrow e}$  & 25.7\%-34.3\% & 24.0\%--33.6\% & 36.8\%--39.0\% \\ \hline
$\mathrm{D^{\pm}\ \rightarrow e}$ & 35.7\%--58.3\% & 39.2\%--56.4\% & 39.7\%--43.0\%  \\ \hline
$\mathrm{D_{s}\ \rightarrow e}$ & 6.7\%--12.9\% & 7.2\%--10.5\% &6.3\%--9.0\%  \\ \hline
$\mathrm{\Lambda_{c}\ \rightarrow e}$ & 3.5\%--10.1\%  & 4.3\%--9.4\% & 2.4\%--14.2\% \\ \hline
$\mathrm{J/\psi\rightarrow e}$ & $<$ 17.0\% & $<$ 15.7\% & $<$ 12.1\% \\ \hline
\end{tabular}}
\end{threeparttable}
\end{table}

\begin{table}
\centering
\caption{Contributions of unsimulated regions of charmed hadron decays ($p_{\mathrm{T}}^{\mathrm{charm}}$ $>$ 20 \GeVc) to the electrons in 0--8 \GeVc.}
\label{tab4}
\begin{threeparttable}
\resizebox{\linewidth}{!}{ 
\begin{tabular}{ccccccc}
& \multicolumn{3}{c}{measured $p_{\mathrm{T}}^{\mathrm{charm}}$ max (\GeVc)} &   \multicolumn{3}{c}{contribution}  \\ \hline
centrality & 0--10\% & 30--50\% & p+p & 0--10\% & 30--50\% & p+p \\  \hline
$\mathrm{D^{0} \rightarrow e}$  & 50 & 35 & 35 & 0.005\% & 0.016\% & 0.026\% \\  \hline
$\mathrm{D^{\pm} \rightarrow e}$ & 50 & 35 & 35 & 0.004\% & 0.012\% & 0.024\% \\ \hline
$\mathrm{D_{s} \rightarrow e}$ & 50 & 50 &25 & 0.037\% & 0.010\% & 0.037\% \\ \hline
$\mathrm{\Lambda_{c} \rightarrow e}$ &24 & 24 &20  & $<$ $10^{-4}$\% & 0.004\% & 0.002\%   \\ \hline
$\mathrm{J/\psi\rightarrow e}$ &  50 & 50 & 40 & $<$ $10^{-4}$\% & 0.006\% & 0.011\%    \\ \hline
\end{tabular}}
\end{threeparttable}
\end{table}

\begin{table*}
\centering
\caption{Uncertainty components of the $\cte$ spectra (0--8 \GeVc) from electron spectra from individual charmed hadron decays.}
\label{tab2}
\begin{threeparttable}
\resizebox{\linewidth}{!}{ 
\begin{tabular}{cccccccccc}
\multirow{3}{*}{} & \multicolumn{3}{c}{from input} &   \multicolumn{3}{c}{from branching ratio} & \multicolumn{3}{c}{from centrality}  \\ \hline
centrality & 0--10\% & 30--50\% & p+p & 0--10\% & 30--50\% & p+p & 0--10\% & 30--50\% & p+p \\  \hline
$\delta\left(\mathrm{D^{0}\ \rightarrow e}\right)$  & 3.5\%--7.6\% & 2.6\%--6.1\% & 1.8\%--4.3\% & 0.4\%--0.6\% & 0.4\%--0.6\% & 0.1\%--0.3\% & 0.5\%--0.7\% & 0.4\%--0.7\% & $\times$ \\  \hline
$\delta\left(\mathrm{D^{\pm}\ \rightarrow e}\right)$ & 5.2\%--39.8\% & 4.0\%--15.7\% & 1.7\%--4.2\% & 0.6\%--1.1\% & 0.7\%--1.1\% & 0.7\%--0.9\% & 0.7\%--1.2\% & 0.7\%--1.2\%  & $\times$  \\ \hline
$\delta\left(\mathrm{D_{s}\ \rightarrow e}\right)$ & 1.1\%--4.8\% & 1.0\%--3.1\% &0.6\%--1.9\% & 0.4\%--0.8\% & 0.4\%--0.7\% & 0.3\%--0.6\% & 0.1\%--0.3\% & 0.1\%--0.3\% & $\times$  \\ \hline
$\delta\left(\mathrm{\Lambda_{c}\ \rightarrow e}\right)$ & 0.8\%--2.6\%  & 1.0\%--2.8\% & 0.3\%--2.1\% & 1.3\%--3.8\% & 1.6\%--3.6\% & 0.9\%--5.4\% & $\times$ & $\times$ & $\times$  \\ \hline
$\delta\left(\mathrm{J/\psi\rightarrow e}\right)$ & $<$ 2.4\% & $<$ 1.8\% & $<$ 0.9\% &$<$ 0.1\% & $<$ 0.1\% & $<$ 0.1\% & $\times$ & $\times$ & $\times$    \\ \hline
\end{tabular}}
\end{threeparttable}
\end{table*}

\begin{figure}
\centering
\includegraphics[width=\linewidth]{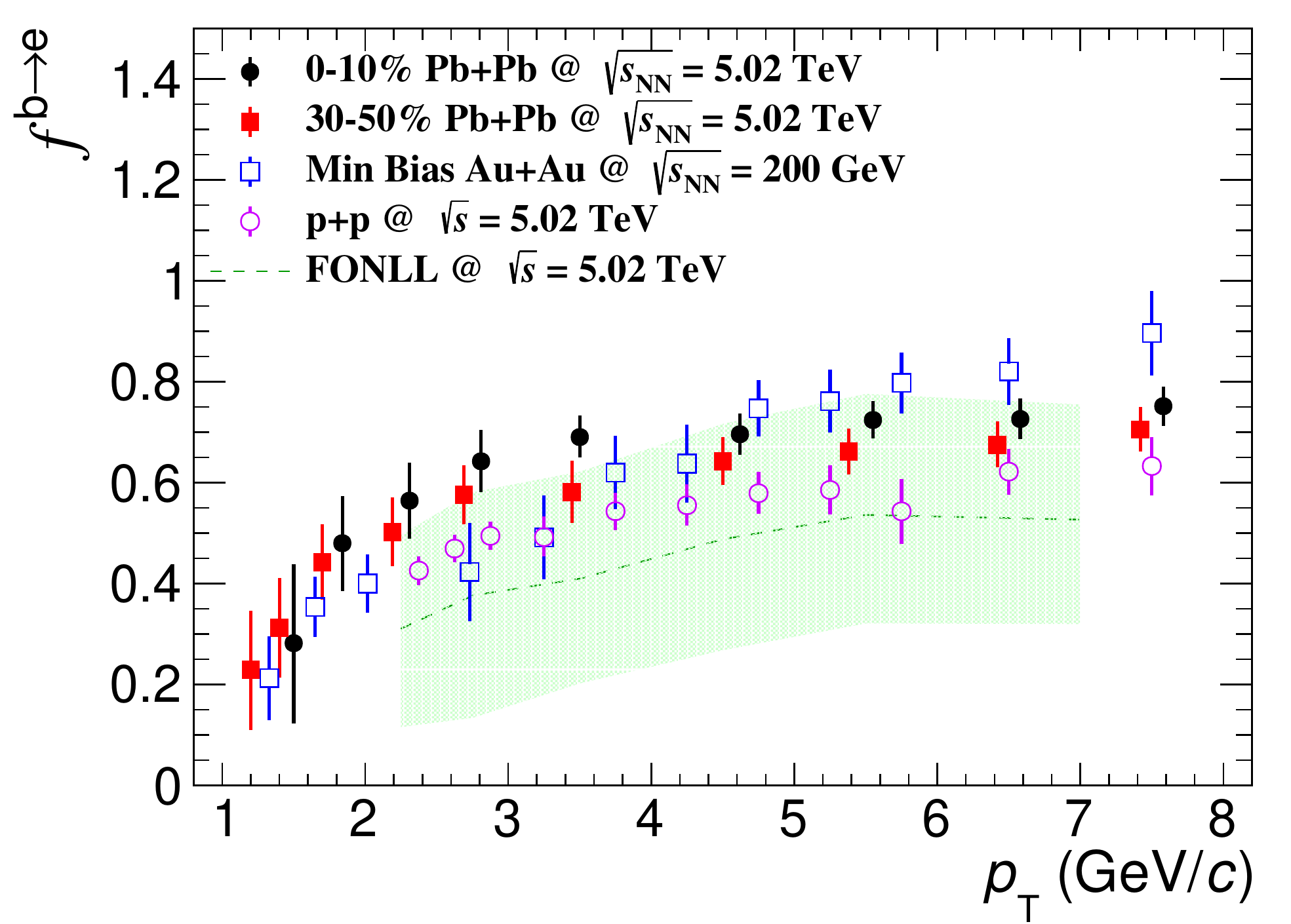}
\vspace{-2ex} 
\caption{Beauty hadron decay electron fractions ($f^{\bte}$) at \sNN = 5.02 TeV in 0--10\% and 30--50\% central Pb+Pb collisions ($\left|\eta\right|$ $<$ 0.8 at \pT\ $<$ 3 \GeVc\ and $\left|\eta\right|$ $<$ 0.6 at \pT\ $>$ 3 \GeVc) and at \s = 5.02 TeV in p+p collisions ($\left|\eta\right|$ $<$ 0.8) compared with those by FONLL calculations~\cite{Cacciari:1998it} and in minimum bias Au+Au collisions at \sNN = 200 GeV ($\left|\eta\right|$ $<$ 0.7)~\cite{Si:2019esg}. }
\label{fig5}
\end{figure}

The beauty contribution fraction in Pb+Pb and in p+p collisions ($\fPb^{\bte}$ and $\fpp^{\bte}$), shown as solid squares (0--10\%), solid circles (30--50\%) and open circles (p+p) in Fig.~\ref{fig5}, is calculated by taking the ratio of the $\bte$ and inclusive HFE spectra, and the \jpsi\ contribution is not removed from the denominator (i.e., the inclusive HFE spectra). The relative uncertainty ($\delta$) of $\fPb^{\bte}$ (5.0\%--55.7\% for 0--10\% and 6.2\%--51.4\% for 30--50\%) and $\fpp^{\bte}$ (5.6\%--11.9\%) is propagated from those of the $\cte$ and the inclusive HFE spectra with 
\begin{equation}
\delta_{f^{\bte}}^2=f_{\mathrm{cb}}^2\left(\delta_{\cte}^2+\delta_{\mathrm{HFE}}^2\right), 
\label{equ1}
\end{equation}
where $f_{\mathrm{cb}}=\left.\left(1-f^{\bte}\right)\middle/f^{\bte}\right.$ in Pb+Pb or p+p collisions. Here we compare the results in 0--10\% and 30--50\% central Pb+Pb collisions, p+p collisions, the previous work in minimum bias Au+Au collisions~\cite{Si:2019esg} and the FONLL prediction~\cite{Cacciari:1998it}. The FONLL prediction of the beauty contribution fraction roughly match the p+p results of this work at \s = 5.02 TeV. The beauty contributions at both centrality classes in Pb+Pb collisions monotonically increase with \pT\ and are significantly higher than that in p+p collisions at \pT\ $>$ 4 \GeVc. At \pT\ $<$ 4 \GeVc, however, similar conclusion is still tenable only in 0--10\% Pb+Pb collisions, since the beauty contribution fractions in p+p collisions are comparable with those in 30--50\% Pb+Pb collisions. At first, for \pT\ $<$ 2 \GeVc, the beauty contributions are smaller than charm, then at \pT\ $\gtrsim$ 3 \GeVc, they overwhelm the charm and start to be flattened, and finally at \pT\ $\sim$ 7 \GeVc, the beauty contributions reach a level of 70\%. It is also observed that the result at 0--10\% centrality are slightly higher than that at 30--50\%, and that the results at Pb+Pb 5.02 TeV show slightly lower than that at Au+Au 200 GeV at \pT\ $>$ 4 \GeVc. This could be due to more energy loss of beauty quark at higher collision energy. However, considering the large uncertainties, the difference is not significant. 

\begin{figure}[!htb]
\centering
\includegraphics[width=\linewidth,height=12.5cm]{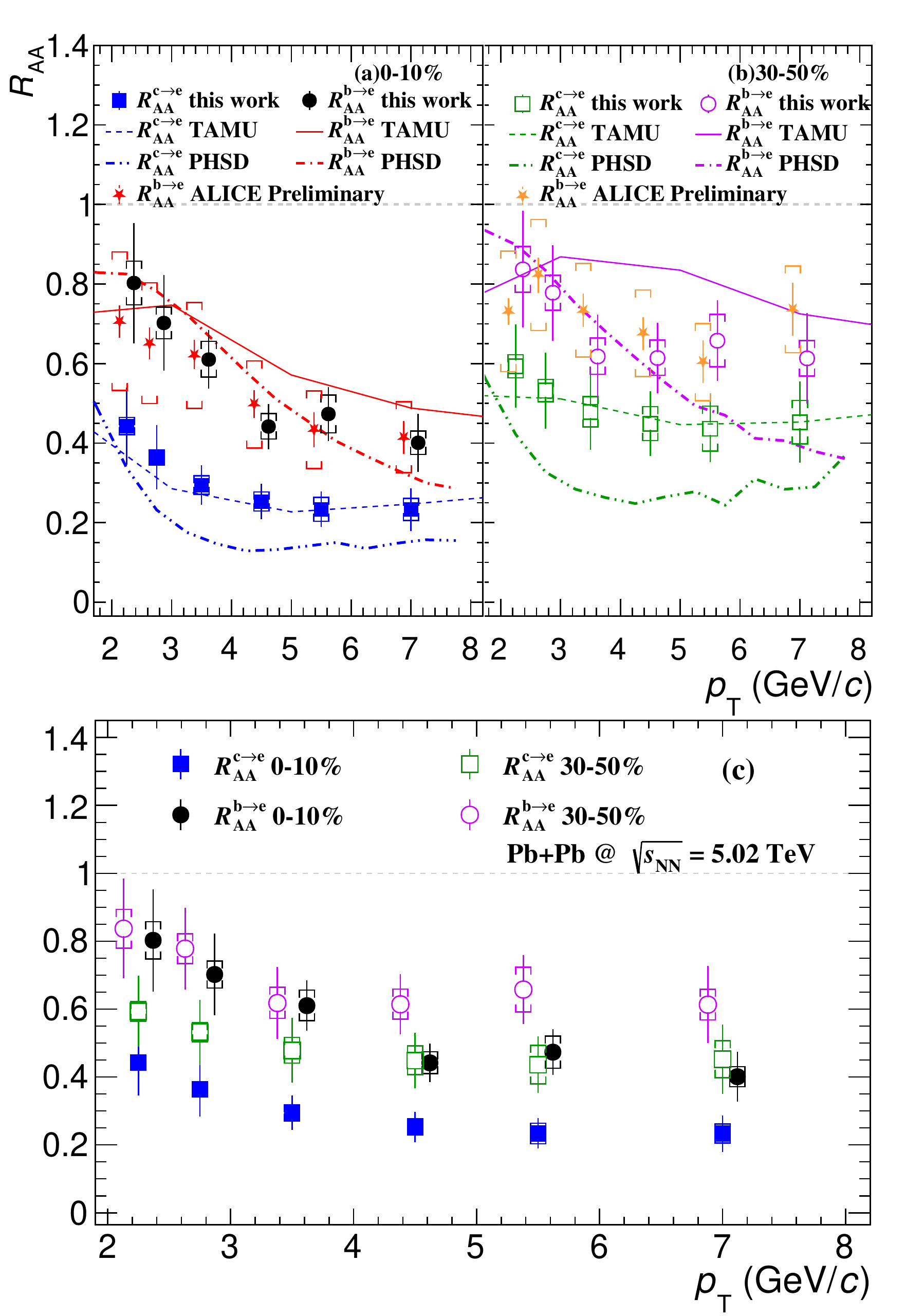}
\vspace{-2ex} 
\caption{The nuclear modification factors ($\RAA$) of $\cte$ and $\bte$ at mid-rapidity ($\left|\eta\right|$ $<$ 0.8 at \pT\ $<$ 3 \GeVc\ and $\left|\eta\right|$ $<$ 0.6 at \pT\ $>$ 3 \GeVc) in Pb+Pb collisions at \sNN = 5.02 TeV. The solid bars represent the total uncertainties in Pb+Pb collisions and brackets denote those in p+p collisions. The ALICE preliminary measurements ($\left|\eta\right|$ $<$ 0.8)~\cite{Park:2021cmg} together with theoretical results by the TAMU model~\cite{Cao:2013ita,Cao:2015hia,Li:2020kax} and by the PHSD model~\cite{Song:2018xca} are shown for comparison with those of this work at 0--10\% (a) and 30--50\% (b) centrality classes. The same results are drawn in panel (c) for the centrality comparison.}
\label{fig6}
\end{figure}

The $\RAA$ of electrons from individual charm and beauty hadron decays ($\RAA^{\cte}$ and $\RAA^{\bte}$) are extracted with the formulae~\cite{Si:2019esg}
\begin{align}
\label{eq1}
\RAA^{\cte}=&\frac{1-\fPb^{\bte}}{1-\fpp^{\bte}}\RAA^{\mathrm{ince}},\\
\RAA^{\bte}=&\frac{\fPb^{\bte}}{\fpp^{\bte}}\RAA^{\mathrm{ince}},
\label{eq2}
\end{align}
where the $\RAA^{\mathrm{ince}}$ represents the $\RAA$ of inclusive electrons from heavy flavor hadron decays ($\left|\eta\right|$ $<$ 0.8 at \pT\ $<$ 3 \GeVc\ and $\left|\eta\right|$ $<$ 0.6 at \pT\ $>$ 3 \GeVc) measured by ALICE~\cite{ALICE:2019nuy}. Figure~\ref{fig6} shows the $\RAA^{\cte}$ and $\RAA^{\bte}$ at 0--10\% (blue solid squares and black solid circles) and 30--50\% (green open squares and violet open circles) centrality classes as functions of \pT\ extracted from Eqs.~\eqref{eq1} and \eqref{eq2}. The results of $\RAA^{\bte}$ are consistent with the ALICE preliminary measurements~\cite{Park:2021cmg} using a different impact parameter method shown as red and orange solid stars at both centrality classes. At each centrality class, clear suppression is observed for both $\RAA^{\cte}$ and $\RAA^{\bte}$ at \pT\ $>$ 2.0 \GeVc, which indicates that charm and beauty quarks strongly interact with the hot-dense medium and lose energy. Meanwhile, $\RAA^{\bte}$ is less suppressed compared with $\RAA^{\cte}$ at \pT\ $>$ 2.0 \GeVc, which is consistent with the mass-dependent energy loss prediction. The $\RAA^{\cte}$ at 30--50\% centrality is observed to be higher than that at 0--10\% at \pT\ $>$ 2.0 \GeVc\ and a similar centrality dependence can be observed for $\RAA^{\bte}$ at \pT\ $>$ 4.5 \GeVc, which suggests that heavy quarks lose more energy in the interaction with the medium created in more central collisions, as seen in Fig.~\ref{fig6}. Theoretical model predictions with the TAMU model~\cite{Cao:2013ita,Cao:2015hia,Li:2020kax} and Parton-Hadron-String Dynamics (PHSD)~\cite{Song:2018xca} are also shown for comparison. TAMU model calculations agree with the $\RAA^{\cte}$ results but seem to overestimate $\RAA^{\bte}$. PHSD model calculations are consistent with $\RAA^{\bte}$ better but underestimate $\RAA^{\cte}$ for both centrality classes. 

From Eqs.~\eqref{eq1} and \eqref{eq2}, the relative uncertainties of $\RAA^{\cte}$ and $\RAA^{\bte}$ can be calculated by
\begin{align}
\label{equ2}
\delta_{\RAA^{\cte}}^2=&f_{\mathrm{bc,Pb}}^2\delta_{\fPb^{\bte}}^2+f_{\mathrm{bc,pp}}^2\delta_{\fpp^{\bte}}^2+\delta_{\RAA^{\mathrm{ince}}}^2,\\
\delta_{\RAA^{\bte}}^2=&\delta_{\fPb^{\bte}}^2+\delta_{\fpp^{\bte}}^2+\delta_{\RAA^{\mathrm{ince}}}^2, 
\label{equ3}
\end{align}
respectively, where $f_{\mathrm{bc}}=\left.f^{\bte}\middle/\left(1-f^{\bte}\right)\right.$ in Pb+Pb or p+p collisions. Table~\ref{tab3} summarizes the uncertainty components of $\RAA^{\cte}$ and $\RAA^{\bte}$.

\begin{table*}
\centering
\caption{Uncertainty components of $\RAA^{\cte}$ and $\RAA^{\bte}$}
\label{tab3}
\resizebox{\linewidth}{!}{
\begin{tabular}{ccccccccc}
\multirow{2}{*}{} & \multicolumn{2}{c}{$\fPb^{\bte}$}  & \multicolumn{2}{c}{$\fpp^{\bte}$} & \multicolumn{2}{c}{$\RAA^{\mathrm{ince}}$} & \multicolumn{2}{c}{total} \\ \hline
centrality & 0--10\% & 30--50\% & 0--10\% & 30--50\% & 0--10\% & 30--50\% & 0--10\% & 30--50\% \\  \hline
$\delta\left(\RAA^{\cte}\right)$ & 13.0\%--17.5\% & 13.1\%--14.9\%  & 5.0\%$-$6.7\% &5.0\%$-$6.7\% & 10.5\%$-$17.4\% & 11.1\%$-$17.4\% & 17.5\%--24.0\% & 18.3\%--23.7\%\\ \hline
$\delta\left(\RAA^{\bte}\right)$ & 5.0\%--13.4\% & 6.4\%--13.7\% & 4.0\%--7.6\%  & 4.0\%--7.6\% & 10.5\%--17.4\% & 11.1\%--17.4\%    & 12.8\%--20.3\% & 14.9\%--19.2\% \\ \hline
\end{tabular}}
\end{table*}

\subsection{Elliptic Flow $v_2$}

Similar procedures to Ref.~\cite{Si:2019esg} are applied to extract the $v_2$ of $\cte$ and $\bte$ and their uncertainties: 

The latest measurement of prompt D-meson $v_2$ in 30--50\% central Pb+Pb collisions at \sNN = 5.02 TeV by ALICE~\cite{ALICE:2017pbx} is parameterized from \pT\ = 0 to 24.0 \GeVc\ with
\begin{equation}
v_2=\frac{p_0n}{1+{\mathrm{exp}}\left(\frac{p_1-\left.\left(m_{\mathrm{T}}-m_0\right)\middle/n\right.}{p_2}\right)}-\frac{p_0n}{1+{\mathrm{exp}}\left(\frac{p_1}{p_2}\right)}+p_3\left(m_{\mathrm{T}}-m_0\right),
\label{eq3}
\end{equation}
where $m_{\mathrm{T}}=\sqrt{p_{\mathrm{T}}^2+m_0^2}$ and $m_0$ denote the transverse and rest masses of the particle, respectively, $n$ is the number of constituent quarks and $p_i$ ($i$ = 0, 1, 2, 3) are free parameters. Meanwhile, the $v_2$ of \lc\ ($n$ = 3) is also obtained from Eq.~\eqref{eq3} by assuming that it follows the NCQ scaling as D-mesons ($n$ = 2). 

The azimuthal angle ($\phi$) distributions of D-mesons and \lc\ in each \pT\ bin is sampled according to the input $v_2$ with
\begin{equation}
\frac{{\mathrm{d}}N}{{\mathrm{d}}\phi}=1+2v_2{\mathrm{cos}}(2\phi).
\label{eq4}
\end{equation}
Then, semileptonic decay simulations are performed with the \pT\ spectra and $\phi$ distributions of D-mesons and \lc\ as inputs. The output $\phi$ distributions of decay electrons are fitted by Eq.~\eqref{eq4} in each \pT\ bin, and the $v_2$ of $\Dte$ ($v_2^{\Dte}$) and $\Lcte$ ($v_2^{\Lcte}$) can be obtained. However, different from the previous work, \jpsi\ also exhibits a finite $v_2$ at LHC energy, so the \jpsi\ contributions (more than 10\% at 3.0 $<$ \pT\ $<$ 4.6 \GeVc\ according to the following calculation) should also be considered. The latest measurement on \jpsi\ $v_2$ only covers the 20--40\% centrality at \sNN = 5.02 TeV~\cite{ALICE:2017quq} and gives poor precision at mid-rapidity ($\left| y\right|$ $<$ 0.9), so the theoretical prediction mentioned in Ref.~\cite{ALICE:2017quq} is applied instead as an estimation~\cite{Du:2015wha,Zhou:2014kka}. Then the $v_2$ of $\jpsite$ ($v_2^{\jpsite}$) is obtained in the same way mentioned above.

The $v_2$ of $\cte$ ($v_2^{\cte}$) and $\bte$ ($v_2^{\bte}$) are calculated with
\begin{align}
\label{eq5}
v_2^{\cte}&=w_{\Dte}v_2^{\Dte}+w_{\Lcte}v_2^{\Lcte}+w_{\jpsite}v_2^{\jpsite},\\
v_2^{\bte}&=\frac{v_2^{\mathrm{ince}}-\left(1-\fPb^{\bte}\right)v_2^{\cte}}{\fPb^{\bte}},
\label{eq6}
\end{align}
where $w_{\Dte}$, $w_{\Lcte}$ and $w_{\jpsite}$ denote the weights by the relative yields of D-mesons, \lc\ and \jpsi\ in $\cte$ calculated in Section~\ref{sec2.1}, and $v_2^{\mathrm{ince}}$ denotes the $v_2$ of inclusive HFE from the parameterized average of the ALICE preliminary measurements ($\left|\eta\right|$ $<$ 0.7) in 30--50\% 5.02 TeV and 20--40\% 2.76 TeV Pb+Pb collisions~\cite{MoreiraDeGodoy:2017wks}. There is no large difference between the $v_2^{\mathrm{ince}}$ measurements at both centrality classes and collision energies, so they are combined together to help perform a better fit and extrapolation.

The statistical and systematic uncertainties of measured D-meson $v_2$ are quadratically summed and propagated into the uncertainties of $v_2^{\Dte}$ and $v_2^{\Lcte}$ through the decay simulations, and they are correlated as a result. The absolute uncertainties ($\sigma$) of $v_2^{\cte}$ and $v_2^{\bte}$ are calculated with
\begin{equation}
\begin{split}
&\sigma_{v_2^{\cte}}=\left( w_{\Dte}\sigma_{v_2^{\Dte}} + w_{\Lcte} \sigma_{v_2^{\Lcte}} \right)^2 + w_{\jpsite}^2\sigma_{v_2^{\jpsite}} ^2\\ 
&+\left( \frac{\Delta v_2\mathrm{[D, \Lambda_c]}}{\cte} \right)^2 \left( w_{\Lcte}^2 \sigma_{\Dte}^2 +w_{\Dte}^2 \sigma_{\Lcte}^2 \right)\\
&+\left( \frac{\Delta v_2\mathrm{[D, J/\psi]}}{\cte} \right)^2 \left( w_{\jpsite}^2 \sigma_{\Dte}^2 +w_{\Dte}^2 \sigma_{\jpsite}^2 \right)\\ 
&+\left( \frac{\Delta v_2\mathrm{[J/\psi,\Lambda_c]}}{\cte} \right)^2 \left( w_{\Lcte}^2 \sigma_{\jpsite}^2 +w_{\jpsite}^2 \sigma_{\Lcte}^2 \right),
\end{split}
\label{eq7}
\end{equation}
where $\Delta v_2\mathrm{[A,B]} = v_2^{\mathrm{A\rightarrow e}}-v_2^{\mathrm{B\rightarrow e}}$, and
\begin{equation}
\begin{split}
\sigma_{v_2^{\mathrm{\bte}}}=&{ f_{\mathrm{1b}}^2\sigma_{v_2^{\mathrm{ince}}}^2 + f_{\mathrm{cb}}^2\sigma_{v_2^{\mathrm{\cte}}}^2}+f_{\mathrm{1b}}^4 \left( v_2^{\mathrm{\cte}}-v_2^{\mathrm{ince}} \right)^2 \sigma_{\fPb^{\mathrm{\bte}}}^2,
\end{split}
\label{eq8}
\end{equation}
where $f_{\mathrm{1b}} = 1/\fPb^{\bte}$ and $f_{\mathrm{cb}} = \left( 1-\fPb^{\bte} \right)/\fPb^{\bte}$. The low values of $\fPb^{\bte}$ at low \pT\ result in large uncertainties of $v_2^{\bte}$.

\begin{figure}
\centering
\includegraphics[width=\linewidth]{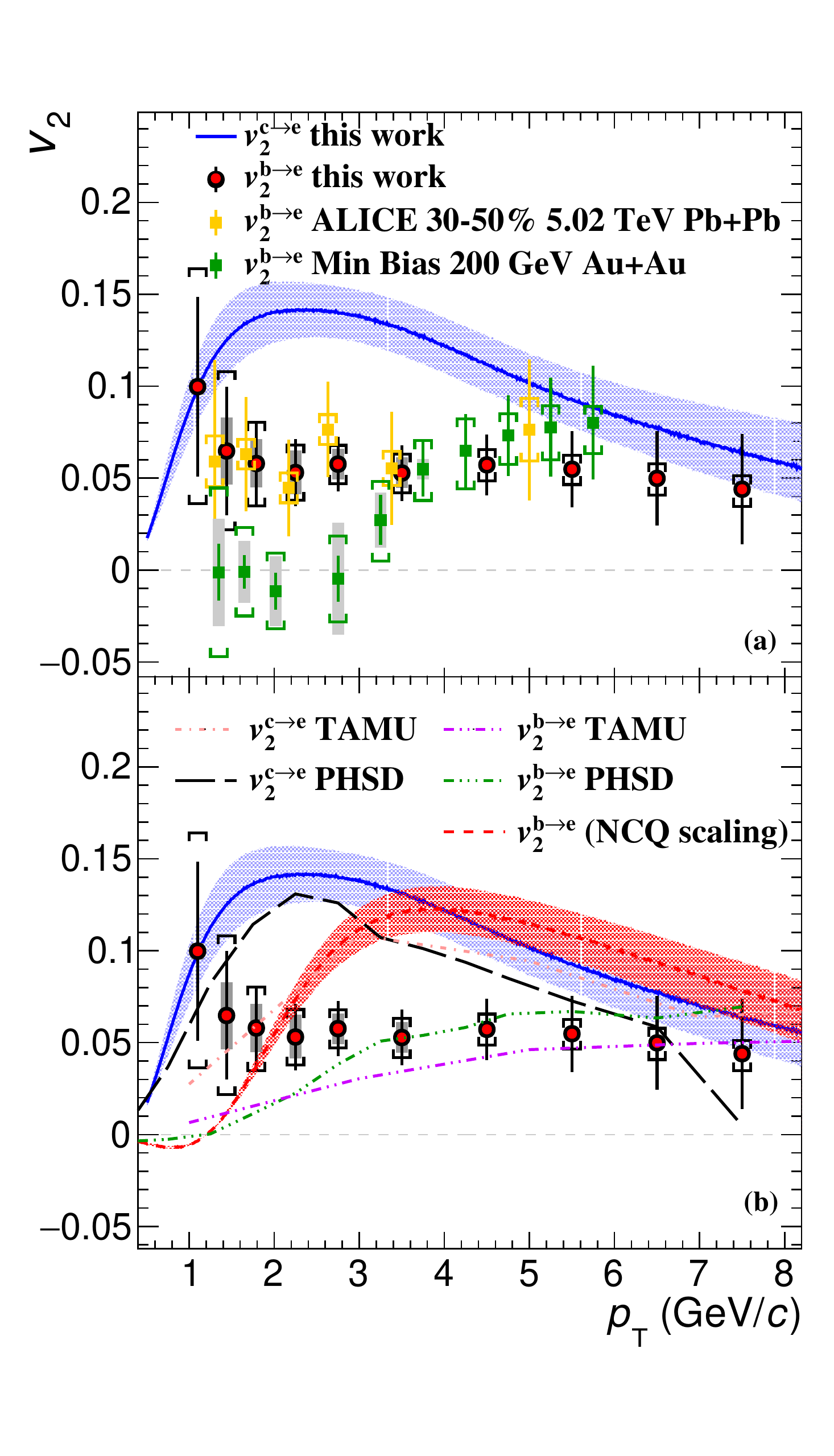}
\vspace{-2ex} 
\caption{The elliptic flows ($v_2$) of $\cte$ and $\bte$ at mid-rapidity ($\left|\eta\right|$ $<$ 0.8) in 30--50\% central Pb+Pb collisions at \sNN = 5.02 TeV. The major contributions to the uncertainty of $v_2^{\bte}$ are from $v_2^{\mathrm{ince}}$ (bars), $v_2^{\cte}$ (brackets) and $\fPb^{\bte}$ (grey bands). In panel(a), the $v_2^{\bte}$ measured in the same collision system by ALICE~\cite{ALICE:2020hdw} and in minimum bias Au+Au collisions at \sNN = 200 GeV~\cite{Si:2019esg} by RHIC are shown as orange and green squares, respectively. In panel(b), the $v_2^{\bte}$ with B-meson $v_2$ NCQ scaling assumption are shown as the red dashed curve. Theoretical predictions of $v_2^{\cte}$ and $v_2^{\bte}$ from TAMU~\cite{Cao:2013ita,Cao:2015hia,Li:2020kax} and PHSD~\cite{Song:2018xca} models are shown as well. }
\label{fig7}
\end{figure}

Figure~\ref{fig7} shows the results of $v_2^{\cte}$ and $v_2^{\bte}$ as the blue solid curve with an uncertainty band and the red solid circles in 30--50\% central Pb+Pb collisions at \sNN = 5.02 TeV , respectively. In Fig.~\ref{fig7}(a), different measurements on $v_2^{\bte}$ are shown for comparison. Measurements by ALICE~\cite{ALICE:2020hdw} are shown as the orange squares with statistical (bars) and systematic (brackets) uncertainties, which are consistent with the results obtained in this work. Results obtained in minimum bias Au+Au collisions at \sNN = 200 GeV~\cite{Si:2019esg} are shown as green squares and are consistent with zero at 1 $<$ \pT\ $<$ 3 \GeVc. While in this work, the $v_2^{\bte}$ results in 5.02 TeV Pb+Pb collisions are observed with a 2.94-$\sigma$ significance ($\chi^2/ndf$ = 17.77/5) deviating from zero and with a confidence level of 85\% ($\chi^2/ndf$ = 6.78/4) to be higher than that in 200 GeV Au+Au collisions, which might be a possible hint of an energy-dependent behavior of beauty quarks reacting with the medium created in heavy-ion collisions. At \pT\ $>$ 3 \GeVc, there is no significant energy dependence seen with current precision.

In Fig.~\ref{fig7}(b), we compare the results of $v_2^{\cte}$ and $v_2^{\bte}$ in this work with those by theoretical predictions. The red dashed curve with an uncertainty band represents the $v_2^{\bte}$ assuming that B-meson $v_2$ follows the NCQ scaling with the same technique as the $v_2^{\Dte}$ calculation and the decay formfactor in the B-meson rest frame sampled by CLEO~\cite{Jain:1995xx} applied. The uncertainty band is calculated from the D-meson $v_2$ uncertainties assuming that with removing the mass effect by $m_{\mathrm{T}}-m_{0}$, the B-meson $v_2$ and D-meson $v_2$ are the same. The $v_2^{\bte}$ of this work deviates from the red dashed curve at 3 $<$ \pT\ $<$ 7 \GeVc\ with a 4.50-$\sigma$ significance ($\chi^2/ndf$ = 31.73/5), and this may indicate that the beauty quark elliptic flow is smaller than that of light and charm quarks which follow the NCQ scaling because the beauty quarks may be too heavy to thermalize and to move with the collective flow of light quarks. The TAMU model predictions~\cite{Cao:2013ita,Cao:2015hia,Li:2020kax} shown as pink dot-dashed curves ($v_2^{\cte}$) and violet dot-dot-dashed curves ($v_2^{\bte}$) are consistent with results of this work at \pT\ $\gtrsim$ 4 \GeVc, while they, especially the $v_2^{\cte}$, deviate from data at low \pT. Another theoretical predictions by PHSD model~\cite{Song:2018xca} are also shown as the black long-dashed curve ($v_2^{\cte}$) and green dot-dot-dot-dashed curve ($v_2^{\bte}$), which agree with data better, but the prediction of $v_2^{\bte}$ still underestimates the data. In a previous work, a compilation of calculations of the dimensionless heavy quark diffusion coefficient, $2\pi T D_s$ that describe the $v_2$ and $\RAA$ of D-mesons, is constrained in the range of 1.5–7 at the critical temperature $T_c$~\cite{ALICE:2017pbx}. In this work, since we only separate the electron contributions from charm and beauty decays, the charm diffusion coefficient extracted from the electron $\RAA$ and $v_2$ from charm decays should be consistent with that extracted from open charm hadrons. The conclusion about the charm diffusion coefficient will not change. However, there is very little discussion on the beauty diffusion coefficient from theoretical calculations due to absence of the measurement on open beauty hadrons in heavy-ion collisions. Usually the heavy flavor spatial diffusion coefficients at zero-momentum reflect the long-wavelength structure of the QCD medium. Therefore, the theoretical treatment of the spatial diffusion coefficients for charm and bottom should be the same. However, there maybe some difference practically between them because the large mass approximation may not work the same way for charm and bottom. Testing the universality of heavy quark spatial diffusion coefficients between charm and bottom is an important next step to constrain the theoretical uncertainties and provide a deep understanding on the QGP microscopic dynamics. Our results could provide a new reference for further theoretical calculations. Especially the electron $\RAA$ and $v_2$ from beauty decays for both RHIC and LHC energies may provide different results for theoretical calculations to extract the QGP transport parameters with different temperatures. 

\section{Summary}

This paper presents an isolation of charm and beauty contributions from heavy flavor decay electrons in p+p and Pb+Pb collisions at $\sqrt{s_{\mathrm{NN}}}$ = 5.02 TeV at LHC. The $p_{\mathrm{T}}$ spectra, $R_{\mathrm{AA}}$ at 0--10\% and 30--50\% centrality classes and $v_2$ at 30--50\% centrality of electrons from charm and beauty decays at mid-rapidity in Pb+Pb collisions at $\sqrt{s_{\mathrm{NN}}}$ = 5.02 TeV are reported. 

The electron $R_{\mathrm{AA}}$ from charm ($R_{\mathrm{AA}}^{\mathrm{c\rightarrow e}}$) and beauty ($R_{\mathrm{AA}}^{\mathrm{b\rightarrow e}}$) decays are suppressed at $p_{\mathrm{T}}$ $>$ 2.0 \GeVc\ in Pb+Pb collisions, respectively, which indicates that charm and beauty quarks interact with the hot-dense medium and lose their energy. The $R_{\mathrm{AA}}^{\mathrm{c\rightarrow e}}$ at 2.0 $<$ $p_{\mathrm{T}}$ $<$ 8.0 GeV/$c$ and $R_{\mathrm{AA}}^{\mathrm{b\rightarrow e}}$ at 4.0 $<$ $p_{\mathrm{T}}$ $<$ 8.0 GeV/$c$ are more suppressed at 0--10\% centrality than at 30--50\% centrality, which suggests that heavy quarks lose more energy in the medium created in more central collisions. However, at low $p_{\mathrm{T}}$ $<$ 4.0 GeV/$c$, $R_{\mathrm{AA}}^{\mathrm{b\rightarrow e}}$ shows no significant centrality dependence. A less suppression of electron $R_{\mathrm{AA}}$ from beauty decays compared with that from charm decays in the whole $p_{\mathrm{T}}$ region from 2.0 to 8.0 GeV/$c$ is observed, which is consistent with the mass-dependent partonic energy loss mechanism. A non-zero electron $v_2$ from beauty decays ($v_{2}^{\mathrm{b\rightarrow e}}$) is observed and consistent with the ALICE measurement in the overlapping $p_{\mathrm{T}}$ region. At low $p_{\mathrm{T}}$ from 1.0 to 3.0 GeV/$c$, a discrepancy between results in 200 GeV Au+Au collisions and in 5.02 TeV Pb+Pb collisions is observed with an 85\% confidence level, which might be a hint that beauty quark reacts differently to the two different media created in the heavy-ion collisions with two very different collision energies. At 3.0 GeV/$c$ $<$ $p_{\mathrm{T}}$ $<$ 7.0 GeV/$c$, $v_{2}^{\mathrm{b\rightarrow e}}$ deviates from a NCQ scaling hypothesis, which favors that beauty quark is unlikely thermalized even in such high energy heavy-ion collisions at LHC. The $R_{\mathrm{AA}}$ and $v_2$ results of electrons with charm and beauty contribution separation could provide a new reference for theoretical models to extract the beauty diffusion coefficient with further theoretical calculations.

\textit{Acknowledgments:} This work was supported by National Natural Science Foundation of China with Grant No.~11890712 and 12061141008, National Key Research and Development Program of China with Grant No.~2018YFE0104700 and 2018YFE0205200, Strategic Priority Research Program of Chinese Academy of Sciences with Grant No.~XDB34000000, and Anhui Provincial Natural Science Foundation with Grant No.~1808085J02.

Dongsheng Li and Fan Si contribute equally to this work as first co-authors. 

\bibliographystyle{cas-model2-names}
\bibliography{references.bib}

\end{document}